\newif\ifsubmode
\submodefalse

\newif\ifprintfig
\printfigtrue

\newif\ifemulate
\emulatefalse

\ifsubmode
  \documentstyle[12pt,aasms4,epsf]{article}
  \received{}
  \accepted{}
  \journalid{}{}
  \articleid{}{}
\else
   \ifemulate
     \documentstyle[emulateapj,epsf]{article}
     \submitted{{\it Submitted for publication in AJ}}
   \else
     \documentstyle[11pt,aaspp4,epsf]{article} 
     \slugcomment{{\it Submitted for publication in AJ}}
   \fi
\fi

\lefthead{van den Bosch et al.}
\righthead{Dark Halos and LSB Rotation Curves}

\newcommand{\etal}{{et al.~}}

\newcommand{\lta}{\lesssim}
\newcommand{\gta}{\gtrsim}

\newcommand{\kmsmpc}{\>{\rm km}\,{\rm s}^{-1}\,{\rm Mpc}^{-1}}
\newcommand{\kms}{\>{\rm km}\,{\rm s}^{-1}}

\newcommand{\Msun}{\>{\rm M_{\odot}}}
\newcommand{\Lsun}{\>{\rm L_{\odot}}}

\begin{document}

\title{Constraints on the Structure of Dark Matter Halos from the 
  Rotation Curves of Low Surface Brightness Galaxies}

\author{Frank C. van den Bosch\altaffilmark{1}, Brant E. Robertson,
and Julianne J. Dalcanton}
\affil{Department of Astronomy, University of Washington, Seattle, 
       WA 98195, USA}

\ifemulate
  \vskip 0.2truecm
\fi

\author{W. J. G. de Blok\altaffilmark{2}}
\affil{Australia Telescope National Facility, Epping NSW 1710, Australia}


\altaffiltext{1}{Hubble Fellow}
\altaffiltext{2}{Bolton Fellow}


\ifsubmode\else
  \ifemulate\else
     \clearpage
  \fi
\fi


\ifsubmode\else
  \ifemulate\else
     \baselineskip=14pt
  \fi
\fi


\begin{abstract}
  We re-examine the disk-halo decompositions of the rotation curves of
  low  surface  brightness (LSB) galaxies with   $V_{\rm  max} \geq 80
  \kms$, taking full account of the effects of beam smearing.  We show
  that the  rotation curves  of these LSB  galaxies are,   contrary to
  previous claims, consistent with dark halos with steep central cusps
  in their density distribution.   In fact, the spatial  resolution of
  the data is not sufficient to put any  meaningful constraints on the
  density profiles of their dark halos, or on cosmological parameters.
  This has  important  implications for  numerous claims made   in the
  literature   regarding  the  halos of  LSB  galaxies,   such  as the
  self-similarity of  their  rotation curves, and their  inconsistency
  with  certain cosmological  models    or   with cold  dark    matter
  altogether.   Only in  one case is   the data of sufficient  spatial
  resolution to obtain   reliable  constraints on  the   slope of  the
  central density  distribution of  the  dark matter  halo.   For this
  single case, we find a  central cusp $\rho \propto r^{-\alpha}$ with
  $0.55 < \alpha < 1.26$ at the  99.73 percent confidence levels. This
  contrasts strongly with the results for  two dwarf galaxies ($V_{\rm
    max} < 70  \kms$) that we analyze, which  yield $\alpha  < 0.5$ at
  the same level of confidence.    This possibly suggests that   halos
  with constant density cores are restricted  to low-mass systems.  We
  show that violent outflows of baryonic matter by supernovae feedback
  can reproduce this mass-dependence of halo cusp slopes.
\end{abstract}


\keywords{dark matter ---
          galaxies: halos ---
          galaxies: fundamental parameters ---
          galaxies: spiral ---
          galaxies: kinematics and dynamics ---
          galaxies: structure}

\ifemulate\else
   \clearpage
\fi


\section{Introduction}
\label{sec:intro}

Cosmological models  make  specific predictions for  the  structure of
dark matter halos.  A generic prediction of the cold dark matter (CDM)
model   is that dark matter  halos  should be  very  dense, with steep
central cusps. Simulations   by Navarro, Frenk  \&  White (1996, 1997)
found a nearly universal density  profile (hereafter the NFW-profile),
which rises as $r^{-1}$ to  the center.  Higher resolution simulations
suggest that  the central profiles should be  even  steeper, rising as
$\rho\!\propto\!r^{-1.5}$ (Fukushige \& Makino 1997, Moore \etal 1998;
Ghigna  \etal 1999), although  other comparable simulations have found
much shallower density cusps ($r^{-0.3}$; Kravtsov \etal 1998, Bullock
\etal 1999).

Observationally, one may hope  to test these predictions by  measuring
the dynamics  of  disk galaxies.   For disks   with a central  surface
brightness close  to  the Freeman (1970)  value,  however, the  actual
insight gained into the structure of dark halos is severely limited by
the    disk's  contribution to  the  circular   velocity.  The unknown
mass-to-light ratio  of  the stellar  component  results  in ambiguity
regarding the actual density distribution of the dark matter. Although
some limits can be obtained (i.e., the maximum  disk), the lack of any
obvious transition  region  from a disk-dominated to  a halo-dominated
part in the  rotation  curves (the ``disk-halo conspiracy'')  severely
impedes the ability to obtain  a unique decomposition.  Many different
combinations   of mass-to-light ratio  and halo  density profile yield
equally good descriptions  of the data  (e.g., van Albada  \etal 1985;
Lake \& Feinswog 1989).

The  structure  of dark  matter halos  is  more directly   revealed in
systems where  the disk contributes  little  to the dynamics.  Several
studies  have noted that the rotation  curves of dark matter dominated
dwarf  galaxies are inconsistent with  steeply  cusped dark halos, and
instead indicate constant density cores (Flores \& Primack 1994; Moore
1994; Burkert  1995;  Navarro, Frenk \&  White 1996;   Burkert \& Silk
1997;   Stil 1999).  This  presents an  important  problem for the CDM
picture, unless as   suggested by Kravtsov  \etal   (1998) and Bullock
\etal (1999),  CDM  halos are  not as  steeply cusped  as suggested by
others.  Several solutions to the problem have already been suggested.
Some argue against CDM, in favor of either modified Newtonian dynamics
(McGaugh \& de Blok 1998a,b)  or self-interacting dark matter (Spergel
\&  Steinhardt 1999).  Others   have  suggested the presence  of  dark
spheroids  of baryons (Burkert \&  Silk 1997),  or violent outflows of
baryonic  matter from stellar feedback  (Navarro,  Eke \& Frenk  1996;
Gelato \& Sommer-Larsen 1999).

An important question in this regard is whether  {\it all} dark matter
halos have  constant density cores as  observed in the dwarfs.  On the
mass   scale of clusters, gravitational  lensing  provides a promising
tool to constrain the density profiles.  So far, however, results have
been contradictory, and the issue is  still under debate (e.g., Tyson,
Kochanski \& Dell'Antonio 1998; Williams, Navarro \& Bartelmann 1999).
On the scale  of massive galaxies,  the most promising  constraints on
the density distributions of dark halos come  from the rotation curves
of low surface brightness (LSB) galaxies.  It has  been shown that the
observed rotation curves  of  LSB galaxies  are remarkably  similar to
those of dwarf galaxies, suggesting similar density profiles for their
dark halos (Kravtsov  \etal 1998; Stil 1999).   Indeed, McGaugh \&  de
Blok (1998a)   have argued that,   as for  nearby dwarf  galaxies, the
rotation curves of LSB galaxies  are inconsistent with the NFW profile
(see also Navarro 1998).

These results imply that constant  density cores are not restricted to
low mass dwarf galaxies.  However, while  the rotation curves of dwarf
galaxies are well resolved (due to  their proximity), the data on more
massive disk galaxies is typically of much  poorer spatial resolution. 
Therefore, before using  low  resolution LSB  rotation curves to   put
stringent constraints  on the nature of  dark matter,  it is important
that the effects of resolution (i.e.\  ``beam smearing'') are properly
taken into account.

In this paper we re-examine a set of HI rotation curves taken from the
literature, focusing on  LSB disk galaxies  with $V_{\rm max} \geq  80
\kms$,  as described in     \S\ref{sec:data}.  We  perform   disk-halo
decompositions by fitting both the observed  HI surface brightness and
rotation curve, including  the effects  of beam  smearing and of   the
adiabatic contraction of the   dark halo.  In \S\ref{sec:modeling}  we
describe these  models,  and  the method  used  to  fit the  data.  We
present the resulting fits for low  resolution and high resolution LSB
data in  \S\ref{sec:lowres} and~\S\ref{sec:highres}, respectively, and
show that  all fits are consistent  with NFW profiles.  We discuss the
implications of  this  result  in \S\ref{sec:disc},  and  summarize in
\S\ref{sec:summ}.

\section{Sample}
\label{sec:data}

To   identify  an appropriate sample  for   measuring  the dark matter
profiles of massive galaxies, we have selected from the literature all
disk galaxies that meet the  following criteria: (i) central  $B$-band
surface brightness fainter than  $\mu_0(B) = 23.0$ mag  arcsec$^{-2}$;
(ii) $V_{\rm max} \geq 80 \kms$; and  (iii) readily available $B$-band
surface  photometry,  HI  rotation curves  and  HI  surface brightness
profiles.   The 16 galaxies  that meet  these  criteria are  listed in
Table~\ref{tab:data}, in order of increasing distance.  In addition we
list two  dwarf galaxies,   DDO~154  and NGC~3109,  which we   use for
comparison in \S~\ref{sec:dwarfs}.

To quantify the resolution of the HI rotation curves, we introduce the
following  two parameters: ${\cal R}_1 \equiv   R_d/S$ and ${\cal R}_2
\equiv R_{\rm max}/S$,  where $R_d$ is  the scalelength of the stellar
disk  (in  the $B$-band),  $R_{\rm max}$  is  the radius   of the last
measured point of the rotation curve, and $S$ is the  FWHM of the beam
of the HI observations  along the beam's  major axis. Larger values of
${\cal R}_1$ and  ${\cal R}_2$ imply  that a rotation curve is  better
spatially resolved and has more independent data points.  The majority
of the galaxies listed in Table~\ref{tab:data} are at relatively large
distances and have ${\cal R}_1  < 1$.  The  only galaxy in the  sample
with $V_{\rm max}  \geq 80 \kms$ that  is well resolved is  the nearby
NGC~247,  which  has  ${\cal R}_1  = 7.4$  and  a  rotation curve that
extends for almost 25 beams.

Throughout this  paper   we adopt a  Hubble   constant  of $H_0  =  70
\kmsmpc$.  All data is  converted to this  value. Where  necessary, we
list  parameter  values  in units  of   $h_{70}$.   All magnitudes and
surface brightnesses  are   corrected  for inclination   and  galactic
extinction.  Because of the  poorly understood dust properties of  LSB
galaxies, we have not attempted a correction for internal extinction.

\section{Modeling}
\label{sec:modeling}

\subsection{Mass components}
\label{sec:components}

When fitting the rotation curves  described above, we consider  models
with three mass-components: an infinitesimally thin  gas disk, a thick
stellar disk, and a spherical dark halo.

Neutral  gas disks in  external galaxies  are typically exponential at
large radii, but often contain a central hole.  To model this profile,
we adopt an HI surface density described by
\begin{equation}
\label{sigHI}
\Sigma_{\rm HI}(R) = \cases{
   \Sigma_0 \, (R/R_{\rm HI})^{\beta} \, {\rm e}^{-R/R_{\rm HI}}  & $(R < R_c)$ ,\cr
   0                                                 & $(R \geq R_c)$ .\cr}
\end{equation}
Here  $R_{\rm  HI}$ is  the scalelength  of  the gas  disk, $R_c$ is a
cut-off  radius, and  $\beta$  describes the magnitude  of the central
suppression in  the HI  distribution.   As we  show below, this  simple
parameterization yields extremely good fits to the observed HI surface
brightness.  The total  HI  mass  that  corresponds  to  this  surface
density is
\begin{equation}
\label{mHI}
M_{\rm HI} = 2 \, \pi \, \Sigma_0 \, R_{\rm HI}^{2} \,
\gamma(\beta + 2, R_c/R_{\rm HI})
\end{equation}
with $\gamma$ the incomplete gamma function.  The circular velocity of
the gas is  computed  using  equation~[2-146] of Binney   \&  Tremaine
(1987), assuming that the disk  is infinitesimally  thin and that  the
total surface density of the gas is  1.3 times that of  the HI to take
account of the mass of helium.
 
For the stellar disk we assume a thick exponential
\begin{equation}
\label{rhostar}
\rho^{*}(R,z) = \rho^{*}_0 {\rm e}^{-R/R_d} {\rm sech}^2(z/z_0)
\end{equation}
where $R_d$ is the scalelength of the  disk.  Throughout we set $z_0 =
R_d/6$. The  exact value of this ratio  does not influence the results
to any significant  degree. The circular  velocity of the stellar disk
is computed using  equation~[A.17] in  Casertano (1983), and  properly
scaled with the stellar  $B$-band  mass-to-light ratio $\Upsilon_B$.   
Except for the two  giant LSB galaxies,  UGC~6614 and F568-6, none  of
the galaxies in our sample has a significant bulge component.
 
We assume that initially  the  dark and  baryonic matter virialize  to
form a spherical halo with a density distribution given by
\begin{equation}
\label{haloprof}
\rho(r) = {\rho_0 \over (r/r_s)^{\alpha} (1 + r/r_s)^{3-\alpha}},
\end{equation}
with $r_s$ being the scale radius of the halo, such that $\rho \propto
r^{-\alpha}$ for  $r \ll  r_s$ and $\rho   \propto r^{-3}$ for $r  \gg
r_s$. For $\alpha =  1$  equation~(\ref{haloprof}) reduces to the  NFW
profile.  We define the   concentration parameter $c =   r_{200}/r_s$,
with $r_{200}$ the radius   inside of which  the  mean density  is 200
times the critical density for closure, i.e.,
\begin{equation}
\label{rvir}
{r_{200} \over h^{-1} {\rm kpc}} = {V_{200} \over \kms}.
\end{equation}
Here $V_{200}$ is the  circular velocity at $r_{200}$, and  $h = H_0 /
100 \kmsmpc$.  The  circular velocity   as function  of radius can  be
written as
\begin{equation}
\label{vhalo}
V_{\rm halo}(r) = V_{200} \sqrt{\mu(xc) \over x \; \mu(c)} 
\end{equation}
with $x = r/r_{200}$ and 
\begin{equation}
\label{fhalo}
\mu(x) = \int\limits_{0}^{x} y^{2 - \alpha} (1 + y)^{\alpha - 3} {\rm d}y
\end{equation}

The initial  collapse    of the gas  and  stellar    disks  within the
virialized halo leads to  a contraction of  the  dark matter halo.  We
assume that the baryons collapse slowly, adiabatically compressing the
halo.  We treat this adiabatic contraction  of the dark halo following
the procedure in Blumenthal \etal (1986) and Flores  \etal (1993).  We
assume that initially the gas follows the same density distribution as
the  dark   matter (equation~[\ref{haloprof}]),  and  assume  that all
baryons end up in  either the gas or the  stellar disk. We thus ignore
any possible hot gas that might be present in  the halo, or that might
have been blown out  of the halo  owing to stellar feedback processes;
given that $V_{\rm max} \geq 80 \kms$ for the  galaxies in our sample,
feedback is not expected to be very efficient (Dekel \& Silk 1986; Mac
Low \& Ferrara 1999).  Our baryon  fraction, defined as $f_{\rm bar} =
(M_{\rm  gas}  + M_{\rm  stars})/M_{200}$,  where  $M_{200}  = r_{200}
V^2_{200} /  G$ is the  total  mass of  the galaxy (baryons  plus dark
matter), thus depends  on the values of  $r_{200}$ and $V_{200}$  that
follow from  fitting the rotation curves, and  can be used as a sanity
check of the best-fit model parameters.

\subsection{Beam smearing}
\label{sec:method}

Because of  the finite resolution of HI  mapping, we need  to convolve
our  model of the   underlying surface brightness profile  $\Sigma(r)$
with the effective point spread  function $P$ (i.e.\  the beam) of the
interferometer, in order to derive the  observed surface brightness at
a position $(x,y)$ on the plane of the sky:
\begin{equation}
\label{convsig}
\tilde{\Sigma}(x,y) = \int\limits_{0}^{\infty} {\rm d}r \, r \, 
  \int\limits_{0}^{2 \pi} {\rm d}\theta \; \Sigma(r') \, 
  P(r,\theta - \theta_0).
\end{equation}
Here $r' = \sqrt{x'^2 + y'^2}$ where $x' = x +  r \cos \theta$ and $y'
= (y + r \sin \theta)  / \cos i$ are the  Cartesian coordinates in the
equatorial plane of the disk, $i$ is the disk's inclination angle, and
$\theta_0$ is the  angle between  the major  axes of the  beam and the
galaxy.  We describe the beam by a two-dimensional Gaussian:
\begin{equation}
\label{beam}
P(r,\theta) = {1 \over 2 \, \pi \, q \, \sigma^2} {\rm exp}\left( -{r^2
\over 2 \sigma^2} \left[ \cos^2 \theta + {\sin^2 \theta \over q^2} 
\right]\right),
\end{equation}
where $q$ is the flattening of the beam, and  $\sigma$ is the Gaussian
dispersion of the beam along its major axis.

Beam smearing  also affects the  observed rotation  curve, by allowing
gas from a wide range of radii  to contribute to the observed rotation
velocity, $\tilde{V}_{\rm rot}$, at a position $(x,y)$ on the plane of
the sky:
\begin{equation}
\label{convvel}
\tilde{V}_{\rm rot}(x,y) = {1 \over \tilde{\Sigma}} 
  \int\limits_{0}^{\infty} {\rm d}r \, r \, 
  \int\limits_{0}^{2 \pi} {\rm d}\theta 
  \; \Sigma(r') \, V_{\rm los}(x',y') \, P(r,\theta - \theta_0)
\end{equation}
(cf. Begeman 1989) where $V_{\rm los}$ is the line  of sight velocity. 
Throughout  we  assume that the gas  moves  on circular  orbits in the
plane of the disk.
 
\subsection{Fitting the data}
\label{sec:fitting}

In order to determine the parameters which best reproduce the observed
surface brightness  distribution $\Sigma_{\rm obs}$ and rotation curve
$V_{\rm obs}$, we adjust our models to minimize
\begin{equation}
\label{chisb}
\chi^2_{\rm HI} = \sum_{i=1}^{N_{\rm HI}}\left( {\Sigma_{\rm obs}(R_i)
    - \tilde{\Sigma}(R_i) \over \Delta \Sigma_{\rm
      obs}(R_i)}\right)^2,
\end{equation}
and
\begin{equation}
\label{chivel}
\chi^2_{\rm vel} = \sum_{i=1}^{N_{\rm vel}}\left( {V_{\rm obs}(R_i) -
    \tilde{V}(R_i) \over \Delta V_{\rm obs}(R_i)}\right)^2.
\end{equation}
Here  $\Delta  \Sigma_{\rm  obs}$ and  $\Delta   V_{\rm obs}$ are  the
errorbars  of   the HI surface   brightness  and  rotation velocities,
respectively, as  quoted in the literature.  $N_{\rm HI}$  and $N_{\rm
  vel}$ are the number of data points.

For a given choice  of the Hubble  constant the models described above
have eight free parameters to fit the  data: $\Sigma_0$, $R_{\rm HI}$,
$\beta$,    $R_c$, $\Upsilon$,  $\alpha$,    $c$,  and  $V_{200}$  (or
equivalently $r_{200}$).  Note, however, that $\Sigma_0$ is completely
determined by normalizing the models to the total  mass in HI, and can
thus be ignored in the fitting routine.  In principle one wants to fit
the  surface  brightness    of  the  gas   and  the    rotation  curve
simultaneously.  However, a  $\chi^2$-minimization analysis using  all
seven  parameters  shows  a very   broad and    noisy minimum and   is
computationally expensive.   We therefore  split the fitting procedure
into two steps.  We start by fitting the  HI surface brightness, after
which the observed rotation velocities  are fit, while holding $R_{\rm
  HI}$,  $\beta$, and $R_c$ fixed at  their best-fit  values.  In both
cases we  use the  downhill-simplex method  (e.g.,  Press \etal  1992)
combined with  a  simple random-walk  routine   to minimize $\chi^2$.  
Tests on model galaxies show that this routine is robust in recovering
the  global  minimum.   Throughout we use   equation~(\ref{rvir}) with
$h_{70}=1$ to set  $r_{200}$, and we require  $c  \geq 1$ and  $0 \leq
\alpha \leq 2$.

In order   to  obtain  confidence levels    around   the best  fitting
kinematical model we proceed as follows.  We set $\alpha$ fixed at its
best fit value and compute $\chi^2_{\rm vel}$ on a $50 \times 50$ grid
of   $(c,\Upsilon_B)$-values  by means   of  a   fast  one-dimensional
$\chi^2$-minimization  routine  to  find  the   best fitting value  of
$V_{200}$. We use  the $\Delta\chi^2$-statistic to  compute confidence
levels on $(c,\Upsilon_B)$ for the  given value of $\alpha$. The  same
procedure  is repeated with either $c$,  or $\Upsilon_B$ held fixed at
their best fitting values.

These confidence levels should not be  interpreted as strict levels of
confidence on  the actual physical parameters.   We have not taken any
uncertainties into account on  the inclination angle, distance, or the
rotation curves of the gas and stellar disks (i.e., we assume that the
stellar  mass-to-light ratio is constant throughout  the galaxy).  Nor
in  most  cases are  the individual   data points  independent of each
other.    Rather, the confidence  levels indicate  to  what extent the
model parameters can be changed {\it  under the given assumptions}, to
obtain fits to the rotation curve that are statistically equivalent.

\section{Low resolution data}
\label{sec:lowres}

In order to  assess the importance  of beam-smearing on low-resolution
data with ${\cal R}_1  \lta 1$ we first analyze  the six galaxies from
the sample   of de Blok,  McGaugh  \& van der   Hulst (1996; hereafter
BMH96) that have $V_{\rm max} \geq 80 \kms$.  We do not analyze F571-8
because its high inclination angle  ($i \simeq 90^{\rm o}$) requires a
somewhat  different modeling technique  than used  here.  The best fit
parameters  to the HI  surface     density  profiles are listed     in
Table~\ref{tab:sbfits}.  The corresponding fits  to the data are shown
in Figure~\ref{fig:sball}.  In each case we obtain an excellent fit to
the data, except for F563-1, where the wiggle  in $\Sigma_{\rm HI}$ at
$\sim 14 h_{70}^{-1}$kpc is not reproduced by our parameterized model.
With the exception of F563-1, which is virtually equally well fit by a
simple exponential,  we find clear  evidence for the presence of large
holes in the HI distributions.

\placefigure{fig:sball}

The  best  fits  to the observed   rotation   velocities are  shown in
Figure~\ref{fig:velall}. The  parameters of   the best-fit models  are
listed in Table~\ref{tab:models}.  Circular  velocities of the stellar
disk are  computed from the $B$-band surface  photometry (de Blok, van
der Hulst \& Bothun 1995).  The best-fit models reproduce the observed
rotation velocities extremely well.  The only exception is F563-1, for
which we fail to reproduce the steep inner rise  of the rotation curve
despite the  steep central cusp of the  best-fit model ($\alpha = 2$). 

\placefigure{fig:velall}

The parameters of the  best fit models span  a wide range. The central
cusp slope   $\alpha$ spans the  entire range   of allowed values from
$\alpha =   2$ (F563-1  and F568-1) to   $\alpha =  0$  (F583-1).  The
mass-to-light ratio, $\Upsilon_B$, spans a range from zero (F563-1 and
F583-1)   up  to  $6.2  \Msun/\Lsun$  (F568-1).   Finally,  the baryon
fraction, $f_{\rm  bar}$, covers  the range   from $0.01$ (F568-3)  to
$0.37$ (F568-1).   Note   that for  realistic  cosmologies  we  expect
$f_{\rm bar}   \simeq 0.05 -   0.1$,  based on  a  baryon  density  of
$\Omega_{\rm bar} = 0.0125  h^{-2}$ (Walker \etal 1991).  Clearly, the
best  fit  models  are  not  always  the  most  physically  meaningful
solutions.  More  interesting, therefore, is to  explore  the areas of
parameter  space that yield   solutions that are  consistent  with the
data.

\placefigure{fig:chia}

Figures~\ref{fig:chia}  and~\ref{fig:chib}   show   contour  plots  of
confidence levels around the best  fit model for three two-dimensional
cuts through parameters space.  In all  cases we find that large areas
of parameter space are consistent with the data. The two extreme cases
are  F568-1   and F574-1 where  virtually   the entire parameter space
probed  falls within the 68.3 percent  confidence level.  Clearly, the
data of  the  BMH96 sample is  of  insufficient spatial  resolution to
meaningfully  constrain the  density distributions of  the dark matter
halos. 

\placefigure{fig:chib}

To  emphasize this point, in  Figure~\ref{fig:mod} we plot the results
for  three models  of  F574-1 that all  fall  within the  68.3 percent
confidence level of the best fit model.  These models are indicated by
small  labeled  dots in Figure~\ref{fig:chib}, and  the corresponding
parameters are listed in  Table~\ref{tab:models}. Models $a$, $b$, and
$c$ have  wildly different mass profiles,  yet  they all fit  the data
remarkably well; the observed   rotation velocities of F574-1  are  of
little use in constraining  the mass model.  

Another   important result is  that  the  maximum mass-to-light ratios
which are consistent with the data are much  larger than those derived
by de Blok \& McGaugh (1997)  using a standard maximal-disk fit (i.e.,
by scaling the contribution  of the disk  to  fit the inner-most  data
point of the rotation curve).  This inner point, however, is the point
most  severely influenced   by   beam smearing,  and   is artificially
reduced, leading  to small mass-to-light ratios (cf., Blais-Ouellette,
Carignan  \& Amram 1998).    Our results indicate  that,  based on the
observed   rotation curves, one can currently   not rule  out that the
central regions  of most of the  LSB galaxies are dominated by baryons
rather than dark matter. 

\placefigure{fig:mod}

In our fitting procedure, we consider a range  of $c$ and $\Upsilon_B$
that  is much  larger than  the  expected spread, leading sometimes to
unphysical  results; for instance,  for small  values  of $c$ we often
find unrealistically high values of  $V_{200}$ (sometimes in excess of
$2000 \kms$).  Clearly, we  do not expect  LSB disks to be embedded in
halos this massive.   In  addition, $B$-band mass-to-light  ratios  of
$\sim 10$ are  rather high given the  typical colors  of LSB galaxies,
and might imply self-gravitating, and hence unstable, disks.  Finally,
we  could be much  more   restrictive  by requiring realistic   baryon
fractions.  However, the  main aim of this  exercise  is to illustrate
that rotation curves with ${\cal R}_1 \lta 1$ {\it themselves} can not
be used to obtain  any meaningful constraints  on the density profiles
or masses of dark halos.

Recently, Swaters (1999)  obtained H$\alpha$ rotation curves for  five
LSB galaxies  in the BMH96  sample (four of  which have  been analyzed
here). These  high resolution data reveal  much steeper inner rotation
rotation curves than for the HI data of BMH96, clearly indicating that
the  HI rotation curves  are severely  affected  by beam smearing. The
H$\alpha$ rotation curves of Swaters (1999) are in good agreement with
our unsmeared best-fit rotation curves.

Reliable and   strict constraints on   the dark  matter  halos of disk
galaxies  thus  requires much higher  spatial  resolution than for the
sample of  BMH96.  Except for  NGC~247, all  galaxies in  Table~1 with
$V_{\rm max}  \geq 80  \kms$ have  been observed with  similar spatial
resolution as the galaxies of the sample  of BMH96 analyzed here.  The
two giant  LSB  galaxies, UGC~6614   and   F568-6, have  ${\cal  R}_1$
somewhat larger than  unity,  but this owes  mainly to   the extremely
large  disk   scalelength.  These   galaxies  have  significant  bulge
components which complicate  constraining the  density distribution of
the halo  component.  We therefore  refrain from analyzing any  of the
other low-resolution data, and shift our attention to the galaxy which
has been observed with the highest spatial resolution: NGC~247.

\section{High resolution data}
\label{sec:highres}

\subsection{The nearby LSB galaxy NGC~247}
\label{sec:ngcgal}

The LSB galaxy  NGC~247 is  a member of  the  nearby ($\sim 2.5$  Mpc)
Sculptor  group and has  been  observed in  HI  by  Carignan \&  Puche
(1998).  Surface photometry   is  taken from  Carignan   (1985b).  The
proximity of NGC~247 ensures that its  HI velocity field has been very
well resolved.  The FWHM of the beam extends only $\sim 13$ percent of
the  scalelength of the  stellar disk,  and the  HI  rotation curve is
measured   out  to 25  beams.   Therefore, this   galaxy is  the ideal
candidate to obtain useful constraints on  the density distribution of
its  dark  matter  halo.   Unfortunately,  to date   it is    also the
{\emph{only}} candidate with  sufficient spatial resolution that  also
meets our selection criteria (see \S~\ref{sec:data}).

Although  for   NGC~247 the effects  of beam   smearing  can easily be
ignored,  we have used  the same analysis  for NGC~247 as  for the low
resolution data.   As  a first  step we fit  the  observed  HI surface
brightness.  No   errorbars for  $\Sigma_{\rm  HI}$ are  given,  so we
assign each data point the same (arbitrary) error in our analysis.  It
turns     out      that    the  particular      parameterization    of
equation~(\ref{sigHI})    yields  a   very   poor  fit.   After   some
experimenting, we found a good fit for a surface brightness profile of
the form:
\begin{equation}
\label{sigHIb}
\Sigma_{\rm HI}(R) = \Sigma_0 \; {\rm exp}\left( R_2 -
 \left[ R_1^{\beta} + R_2^{\beta} \right]^{1/\beta}\right),
\end{equation}
with $R_1  = R/R_{\rm HI}$ and  $R_2 = R_c/R_{\rm  HI}$.  The best fit
parameters are listed in Table~\ref{tab:sbfits}, and the resulting fit
is shown in the upper left panel of Figure~\ref{fig:highres}.

\placefigure{fig:highres}

The upper   right  panel shows  the  best  fit model  to  the observed
rotation curve.  This  model is completely  dark matter dominated, and
has  a dark halo with  $\alpha = 1.02$,   $c=7.2$, and $V_{200} = 93.1
\kms$.   The  best-fit  mass-to-light  ratio of  the   stellar disk is
$\Upsilon_B   = 1.0 \Msun/\Lsun$, resulting in    a baryon fraction of
$f_{\rm   bar}  =  0.011$.   Figure~\ref{fig:chic}  shows  that  these
parameters are well constrained. We can rule out models with $\alpha <
0.55$ and  $\alpha > 1.26$ at the  99.73 percent confidence level; the
improvement with   respect to the    low resolution data of  BMH96  is
apparent.   Note   that the parameters   of  the  best  fit model  are
remarkably    consistent  with a NFW    profile   with a concentration
parameter that is typical for realistic cosmologies (Navarro, Frenk \&
White 1997).

\subsection{Comparison to nearby dwarfs}
\label{sec:dwarfs}

In order to  contrast the results for NGC~247  with dwarf galaxies, we
have  applied   the  same analysis  to    DDO~154  and NGC~3109. These
galaxies,  observed by  Carignan   \&  Beaulieu  (1989) and Jobin   \&
Carignan  (1990)   respectively,  were shown   by  Moore  (1994) to be
inconsistent with a  dark matter halo  that follows a Hernquist (1990)
profile.  Both galaxies are nearby  and have  been observed with  good
spatial resolution  (see  Table~\ref{tab:data}).  For NGC~3109  ${\cal
  R}_1 = 5.2$ and ${\cal R}_2 = 27.8$ such that it has one of the best
resolved   HI velocity  fields.    While    DDO~154 has only    ${\cal
  R}_1\sim0.7$, the radius  where the rotation  curve changes slope is
significantly larger than the beam size.

\placefigure{fig:chic}

Our best fit models to the data  are shown in Figure~\ref{fig:highres}
and listed  in Tables~\ref{tab:sbfits}  and~\ref{tab:models}.  As  for
NGC~247, no errorbars for the HI surface density are available, and we
assign  each  data point  the  same (arbitrary)  error.   The best fit
kinematic models of both  galaxies are remarkably similar.   Both have
$\alpha  =  0$  ($\alpha \lta  0.5$  at  the 99.73  percent confidence
level), $c \approx 12$ and $\Upsilon_B = 0$.  The contour plots of the
confidence  levels are  shown in  Figure~\ref{fig:chic}.  Clearly, the
observed  rotation curves  of    these  dwarfs constrain the    models
extremely well, although the  unphysical mass-to-light  ratio suggests
that our  underlying  model for either  the surface  brightness or the
detailed   halo shape  must    not be quite    correct.  Although firm
conclusions cannot be made on the basis of  three galaxies only, these
preliminary   results suggest  that constant density    cores may be a
feature of only the least massive galaxies. The low resolution data of
BMH96 is at least consistent with this picture.

\section{Discussion}
\label{sec:disc}

The disk-halo decompositions of  LSB  galaxies with $V_{\rm max}  \gta
80\kms$ presented in \S\ref{sec:lowres} call into question a number of
strong claims  that have been made in  the literature. Foremost is the
conclusion that  the  HI rotation  curves   of LSB disk  galaxies  are
inconsistent with dark matter  halos with steep central cusps (McGaugh
\& de  Blok 1998a;  Navarro  1998;  Salucci \&   Persic 1997),  and by
implication, that observations are  inconsistent with the  predictions
of CDM cosmologies (Navarro, Frenk  \& White 1997; Fukushige \& Makino
1997; Moore \etal 1998).  In contrast  to previous interpretations, we
find  that for  all but  the lowest  mass  galaxies ($V_{\rm max} \lta
60\kms$) the observations are of sufficiently low resolution that they
place  little or  no constraints on   the inner shapes  of dark matter
density profiles.  In fact, the single case which we have found in the
literature   which is capable    of limiting the   dark matter profile
(NGC~247;  \S\ref{sec:highres})   has   an  inner     slope which   is
{\emph{exactly}} in  agreement  with the  CDM  predictions of Navarro,
Frenk \& White (1997).

Recently,  Kravtsov \etal (1998) claimed   that the rotation curves of
LSB galaxies  are self-similar and  have the same   shape as for dwarf
galaxies (see  also Burkert 1995).   This presents another problem for
CDM, since the  actual density profiles of dark  halos (and  hence the
corresponding rotation curves) are  expected  to reveal a  significant
amount of  scatter (Avila-Reese \etal 1999;  Ying 1999;  Bullock \etal
1999).  However, our results suggest that the observed self-similarity
is likely to be a consequence of the large  amount of beam smearing in
the existing LSB observations.

The apparent shallowness of LSB   rotation curves had previously  lead
astronomers to place strong upper limits  on $\Omega_0$.  For example,
using halo-only NFW fits to a wide variety of disk galaxies, Pickering
\etal (1997)  and   Navarro (1998) derived  low  upper  limits for the
concentration of the halo, and thus  argued for a low-density Universe
with  $\Omega_0 \lta 0.2$, with  the strongest constraints coming from
LSB galaxies from the samples of BMH96 and Pickering \etal (1997).  As
the fits to these  samples did not  include beam smearing, the results
presented here suggest that the  best-fit halo concentration is likely
to have  been severely underestimated,  and thus much larger values of
$\Omega_0$  are permitted  by the  data.    For comparison, our   most
restrictive case in our  analysis, F568-3, has  $c < 8.3$ at  the 68.3
percent confidence level; this  limit would be  even higher if  we had
not taken account  of the contributions of the  gas and stellar disks,
and adiabatic contraction of the halo.

While  our   analysis suggests  that  the   halos of  massive galaxies
($V_{\rm  max} \geq    80\kms$) are fully   consistent  with centrally
concentrated  dark matter  halos   and  thus  with  large   values  of
$\Omega_0$, we find that the halos of  dwarf galaxies continue to tell
a different story.  Unlike their  more massive LSB counterparts, there
is compelling evidence that the halos of dwarf galaxies do indeed have
shallow central cores.  If  future high-resolution studies  reveal the
existence of large cores in massive LSB galaxies as well, we will need
to revise the  standard  model for the nature  of  dark  matter (e.g.,
Spergel \& Steinhardt 1999).   If, however, these studies confirm  the
presence  of steep central cusps in  galaxies other than NGC 247, then
the data on  dwarf galaxies suggest that there  is an intrinsic change
in the  structure of  dark matter  halos  at $V_{200} \sim   100\kms$. 
Although this violates the scale-free behavior predicted by almost any
variant of collisionless dark matter models,  it is not necessarily in
contradiction with these models, if, for instance, it can be caused by
baryonic  physical processes.  One  of   the most promising  candidate
processes is outflows (or ``feedback''),  a process which is known  to
exist  in dwarf galaxies  (e.g., Puche  \etal  1992; Hunter, Hawley \&
Gallagher  1993; Martin 1999),  and which is  necessary to explain the
slope  and scatter of  the Tully-Fisher relation  (e.g., van den Bosch
1999).  After outflows remove mass from the  center of the galaxy on a
short timescale, the remaining   material revirializes to form  a less
centrally   condensed system.   $N$-body  simulations  of this process
(Navarro  \etal  1996) have shown  that  violent outflows  of baryonic
matter from the center of a dark  halo do result  in the creation of a
constant density core (see also Gelato \& Sommer-Larsen 1999).

We wish (1)  to test if outflows  can be sufficiently large to produce
cores in   dwarf galaxies, and  (2)  to measure the velocity  scale at
which this process   becomes ineffective.  In the  Appendix,  we use a
simple prescription  to   estimate how  the  structure  of the  galaxy
responds   to a realistic outflow.   We  assume that supernovae expel
some fraction of  the baryonic  mass from the  galaxy  on a time-scale
that  is short compared to the   dynamical time (i.e.,  resulting in a
non-adiabatic  change  of the  potential),  and we  tune the  feedback
efficiency  to match  the  Tully-Fisher   relation.  The blow-out    of
baryonic material leads to a change in the profile  of dark matter. We
assume that the mass of dark matter which can essentially be displaced
by this process is proportional to the mass  of baryonic material lost
by  the galaxy, with some   proportionality constant $\eta$.  Assuming
that when  the halo revirializes,  the ``rearranged'' dark matter mass
winds up outside of the scale radius $r_s$, one  can compute the final
cusp slope $\alpha_f$ of the revirialized halo.

\placefigure{fig:alpha}

The resulting cusp   slopes are plotted  in Figure~\ref{fig:alpha}  as
function of $V_{200}$. The locations of  NGC~247, DDO~154 and NGC~3109
are plotted for  comparison. Results are shown  for  halos with $c=5$,
10, and 20, which roughly covers the range of expected values, and for
initial halos with  both $\alpha_i =  1$ (NFW profile) and $\alpha_i =
1.5$ (the  value suggested by Moore  \etal 1998).   The models clearly
show  that the final cusp  slope decreases  with decreasing halo mass,
such that dwarf  galaxies could have  significantly flatter cusps than
more massive galaxies, in spite of having the same initial dark matter
profile. If  $\eta \simeq 1$, halos  initially follow  an NFW profile,
and have $c \simeq 10$  (as for NGC~3109,  DDO~154, and NGC~247), this
model predicts constant density cores in systems with $V_{200} \lta 75
\kms$,  in reasonable  agreement    with observations.   If,  however,
$\alpha_i = 1.5$, there  is too much mass  in  the center for  violent
outflows to destroy the central cusp;  even systems with $V_{200} = 30
\kms$  will not  have  constant  density  cores.   In this case,   the
efficiency of relocating  dark matter   has  to be larger. For   $\eta
\simeq 2$, the  resulting cusp slopes are in  good  agreement with the
high-resolution  data analyzed here.   These results suggest that even
if  the initial structure  of dark matter halos   is identical for all
galaxies, ``gastrophysics'' can  easily introduce a systematic  change
in the profile with mass.

While this  adhoc procedure  ignores the  detailed  physics of how the
dark matter adjusts to an instantaneous mass loss,  it is still useful
for setting  the overall scale over which  outflows could be effective
in   creating dark matter  cores.   Detailed  $N$-body simulations are
required  to test  the importance of   our assumptions (i.e., constant
$r_s$ and $c$) and  to obtain estimates  of  $\eta$. Our results  also
depend on     the    particular model    for     the feedback   (i.e.,
$\epsilon^0_{\rm SN}$ and $\nu$, see the Appendix). With these caveats
in mind,  it is nevertheless reassuring that  this  simple model, with
feedback    parameters that are  motivated by   empirical  data on the
Tully-Fisher relation, yields results that are at least of the correct
order of  magnitude to reconcile the constant  density cores in dwarfs
with CDM predictions.

\section{Summary and Conclusions}
\label{sec:summ}

We  have re-examined the disk-halo decompositions  of several LSB disk
galaxies  with  $V_{\rm max} \geq   80 \kms$, improving  upon previous
studies  by   taking beam  smearing   and  adiabatic contraction  into
account.  Contrary to previous claims we find that the rotation curves
of these systems  are consistent  with dark  halos with  steep density
cusps.   The actual  slope   of  this  density  cusp  is   very poorly
constrained: the data can not be used to argue against either constant
density cores  or against very steep cusps  such as those suggested by
Fukushige \&  Makino (1997) and  Moore \etal (1998).  Our results also
call into question the observed self-similarity of LSB rotation curves
(Kravtsov   \etal   1998), and  strongly    reduce  the constraints on
$\Omega_0$ (Pickering \etal 1997; Navarro 1998).

While we have shown that existing HI rotation  curves for LSB galaxies
are  consistent  with   predictions of  CDM  and   a  universal baryon
fraction, they cannot  place any meaningful constraints.  Observations
at much  higher resolution are  required.  Given the typical beam size
of  HI synthesis observations, this   limits  us to relatively  nearby
galaxies.  Most nearby LSB galaxies,  however, are dwarf galaxies with
$V_{\rm max}  \lta  60  \kms$.  The  only  LSB  galaxy that  meets our
selection criteria and  is well resolved is NGC~247,  for which the HI
rotation curve extends  for almost 25 beams,  and constrains the model
parameters very well  ($0.55  \leq \alpha   \leq  1.26$ at  the  99.73
percent confidence level with  a  best-fit cusp   slope of $\alpha   =
1.02$).   This  contrasts strongly  with  the  nearby dwarf  galaxies
DDO~154 and NGC~3109, where we find $\alpha \leq 0.5$ at 99.73 percent
confidence with a  best-fit value of  $\alpha =  0$ (i.e., a  constant
density core).  This  suggests that  constant density cores  are  only
present in dwarf  galaxies,  whereas more massive galaxies  have  dark
halos  with a steep central cusp  as expected for   CDM. We have shown
that violent outflows due to supernovae  feedback can account for such
a mass-dependence of the shape of dark matter halos.

Although the low-resolution data examined here is consistent with this
picture, it  is    premature to draw  conclusions   about  the density
profiles at high mass based upon the single case of NGC~247.  What are
needed at    this point are high  resolution   rotation curves   for a
sufficiently large sample of LSB disk galaxies  with $V_{\rm max} \gta
80 \kms$. Unless these galaxies are relatively nearby ($\lta 10$ Mpc),
there  is  little hope  that HI   observations will  yield  sufficient
spatial   resolution.  We need  to   focus  on observations in  either
H$\alpha$  (i.e., Blais-Ouellette  \etal  1998;  Swaters 1999)  or  CO
(Sofue \etal  1999).   These  tracers of  the   velocity field  can be
observed at much higher spatial resolution, and, when combined with HI
rotation  curves  to sample  the outer parts   of  the velocity field,
provide promising tools to examine  whether more massive disk galaxies
have cusped dark halos similar to NGC~247. Independent of the outcome,
such  studies will  have strong  implications  for theories of  galaxy
formation and for the nature of the dark matter.


\acknowledgments

We are grateful   to Craig Hogan for  a  critical reading of an  early
draft of the paper, and to George Lake and  Julio Navarro for valuable
discussions.   FvdB was supported  by  NASA through  Hubble Fellowship
grant \#  HF-01102.11-97.A  awarded  by the  Space   Telescope Science
Institute, which is  operated   by AURA  for NASA under   contract NAS
5-26555.  BER was supported by the  Washington Space Grant Consortium,
under a grant from NASA, and by the Mary Gates Endowment for Students.


\clearpage

\begin{appendix}

\section{Supernovae feedback and the mass-dependence of halo cusp slopes}
\label{sec:fb}

We examine if outflows can explain a change  in the central cusp slope
of dark matter  halos from $\alpha  \simeq 0$ for systems with $V_{\rm
  max} \lta 70 \kms$ (i.e., NGC~3109) to $\alpha \simeq 1$ for systems
with $V_{\rm max} \gta 100 \kms$ (i.e., NGC~247).

As a  mechanism for producing  outflows of baryonic matter we consider
feedback from supernovae (SN).  The total energy produced by SN can be
written as
\begin{equation}
\label{enerSN}
E_{\rm tot} = \eta_{\rm SN} \, E_{\rm SN} \, M_{*} 
\end{equation}
with $\eta_{\rm SN}$ the number of SN produced per Solar mass of stars
formed, $E_{\rm SN}$ the energy produced per SN, and $M_{*}$ the total
mass in stars.  We assume that this energy  expels a mass $M_{\rm fb}$
of baryons  from the  halo.   Taking  account of  the system's  escape
velocity and requiring energy balance, one obtains
\begin{equation}
\label{mblowout}
M_{\rm fb} = {\epsilon_{\rm SN} \, \eta_{\rm SN} \, E_{\rm SN} \,
M_{*} \over V_{200}^2},
\end{equation}
(cf.   Kauffmann \etal 1993;  Natarajan  1999) where we introduce  the
parameter  $\epsilon_{\rm SN}$  which  describes  the efficiency  with
which the  SN energy expels  baryonic mass from the halo. Conservation
of  baryon  mass\footnote{Note  that   we  make   the  oversimplifying
  assumption that all the baryonic matter is  in stars or expelled; no
  cold gas is taken into account.} requires that $M_{*} + M_{\rm fb} =
f_{\rm bar} \, M_{200}$.  Setting  $\eta_{\rm  SN} = 4 \times  10^{-3}
\Msun^{-1}$, $E_{\rm SN} = 10^{51}$ ergs, and writing
\begin{equation}
\label{epsSN}
\epsilon_{\rm SN} = \epsilon^0_{\rm SN} \, \left({V_{200} \over 250
\kms}\right)^{\nu}
\end{equation}
one obtains
\begin{equation}
\label{mblow}
M_{\rm fb} = {f_{\rm bar} \, M_{200} \over 1 + {1 \over 3.22 \,
\epsilon^0_{\rm SN}} \left({V_{200} \over 250 \kms}\right)^{2-\nu}}.
\end{equation}
As  a model for the feedback  we consider $\epsilon^0_{\rm SN} = 0.05$
and $\nu =  -0.3$ and we  set the  baryon  fraction to $f_{\rm bar}  =
0.085$.  Semi-analytical models  for the  formation of disk  galaxies,
presented  by van  den Bosch (1999)  and  van  den Bosch \&  Dalcanton
(1999) have  shown that for  these values one  obtains a near-infrared
Tully-Fisher    relation   with   the   correct   slope,  scatter  and
normalization.   In addition, adopting    a feedback model  with these
parameters  yields an absence    of   high surface  brightness   dwarf
galaxies, as observed.

We now assume  that the blow-out  of mass $M_{\rm  fb}$ results in  the
relocation of  a dark matter mass of  $\Delta M_{\rm DM} = \eta M_{\rm
  fb}$ to larger radii,  with  $\eta$  of order  unity.  In  order  to
quantify $\Delta M_{\rm DM}$ in terms  of the cusp  slopes of the dark
halos, we  make the following  simplifying assumption.  We assume that
both the   initial and final   halos have  density  profiles given  by
equation~(\ref{haloprof}) with   $\alpha  =  \alpha_i$  and  $\alpha =
\alpha_f$,  respectively. In addition, we  assume that both halos have
identical dark  matter masses,  scale-radii, $r_s$,  and concentration
parameters, $c$.   Defining $\Delta M_{\rm DM}$  as  the difference in
halo mass within a radius $r_s$ we obtain
\begin{eqnarray}
\label{delta_mass}
\lefteqn{\Delta M_{\rm DM} \equiv M_i(r_s) - M_f(r_s)}
\;\;\;\;\;\;\;\;\;\;\; \nonumber \\
& = (1 - f_{\rm bar}) \, M_{200} \, \left[ {\mu_i(1) \over \mu_i(c)}
- {\mu_f(1) \over \mu_f(c)} \right] 
\end{eqnarray}
with $\mu_i(x)$  and $\mu_f(x)$ given  by equation~(\ref{fhalo})  with
$\alpha$ set to the initial and final values, respectively.

Using  equations~(\ref{mblowout})  and~(\ref{delta_mass})    one   can
compute  the  cusp slope  $\alpha_f$  as function  of $V_{200}$,  $c$,
$\alpha_i$, and $\eta$ by solving for the root of $\Delta M_{\rm DM} -
\eta M_{\rm fb} = 0$.  The results are shown in Figure~\ref{fig:alpha}
and discussed in \S~\ref{sec:disc}.

\end{appendix}


\ifemulate\else
  \baselineskip=10pt
\fi


\clearpage


\ifsubmode\else
\baselineskip=14pt
\fi


\newcommand{\figcapsball}{Fits    to the   observed  (uncorrected  for
  inclination) HI surface brightness (open circles with errorbars) for
  six galaxies of the sample of BMH96.  The thick  solid lines are the
  best   fits  to the   data   (see  Table~\ref{tab:sbfits}   for  the
  parameters), and the thin  dashed lines correspond to the  unsmeared
  face-on HI surface brightness profile.  The downward arrow indicates
  the FWHM  of  the beam along   its major axis,   and is  plotted for
  comparison.  Note  that in five  out of  six  cases the unsmeared HI
  surface brightness  profiles reveal  clear   evidence for a  central
  hole. \label{fig:sball}}

\newcommand{\figcapvelall}{Fits    to  the   observed  (corrected  for
  inclination) rotation velocities  (open circles with errorbars)  for
  six  galaxies of   the  sample of  BMH96.    The  thick solid  lines
  correspond  to the best   fitting models (see Table~\ref{tab:models}
  for  the parameters).  In addition,  we  plot the unsmeared circular
  velocity   curves  of   the models  (thin   solid   lines),  and the
  (unsmeared) contributions  from the  halo (short-dashed  lines), the
  stellar disk (long-dashed lines), and  the gas disk (dotted lines).  
  If  no long-dashed line  is plotted, it  means that the best fitting
  model has zero mass-to-light ratio. \label{fig:velall}}

\newcommand{\figcapchia}{Grayscale and   contour plots   of confidence
  levels around the  best fit models  for F563-1, F568-1,  and F568-3. 
  For  each galaxy three different  cuts  through parameter space  are
  shown:    the $({\rm  log}  c,   \alpha)$-plane  (left panels),  the
  $(\Upsilon_B, \alpha)$-plane (middle panels), and the $({\rm log} c,
  \Upsilon_B)$-plane (right panels). The  big white dot  indicates the
  best-fit  model.  Contours of only three   levels  of confidence are
  shown:  68.3 percent (thickest   contour), 95.4 percent  (contour of
  medium     thickness),      and        99.73    percent    (thinnest
  contour).\label{fig:chia}}

\newcommand{\figcapchib}{Same    as  Figure~\ref{fig:chia}  but    for
  F568-V1,  F574-1, and F583-1. For  F574-1 three small dots, labeled
  $a$, $b$, and $c$ are plotted.  These correspond to the locations of
  three     models   discussed     in    the     text    (see     also
  Figure~\ref{fig:mod}).\label{fig:chib}}

\newcommand{\figcapmodels}{Fits to  the rotation curves  of F574-1  of
  three  different  models that  all   fall  within the  68.3  percent
  confidence   level    of    the  best-fit      model  presented   in
  Figure~\ref{fig:velall}. The labeling  of  the different curves  is
  the same as in that figure. The parameters of the various models are
  listed  in  Table~\ref{tab:models}.  Note  that models  a, b, and c,
  which  all  have    wildly different parameters,   yield   virtually
  indistinguishable   rotation   curves,   even though  the  unsmeared
  circular  velocity curves  (thin  solid lines)  are  very different. 
  This illustrates   the  large amount    of   freedom in   the  model
  parameters,   which   is also apparent   from  the  contour plots in
  Figure~\ref{fig:chib}.\label{fig:mod}}

\newcommand{\figcaphighres}{Best  fits  to  the  HI surface brightness
  (left panels) and HI rotation curves (right panels) for the galaxies
  in Table~\ref{tab:data}  that have been   observed with high spatial
  resolution: NGC~247, DDO~154,  and NGC3109.   Symbols and lines  are
  the  same as  in Figures~\ref{fig:sball} and~\ref{fig:velall}. Since
  we  have  adopted distances to  these nearby  galaxies that have not
  been  determined from  the   recession  velocities, the   radii  are
  independent of $h_{70}$.\label{fig:highres}}

\newcommand{\figcapchic}{Same      as   Figures~\ref{fig:chia}     and
  \ref{fig:chib} but for NGC~247,  DDO~154, and NGC~3109. Note that we
  now  only  plot  results for   $\Upsilon_B \leq   2.5  \Msun/\Lsun$. 
  \label{fig:chic}}

\newcommand{\figcapfb}{The  final  cusp slope, $\alpha_f$, as function
  of $V_{200}$, depicting the effect of SN  feedback on the cusp slope
  of   dark  matter halos.  Results   are  plotted for three different
  values of $c$, as  labeled in the  upper right panel. Panels on  the
  left  correspond to  halos with an   initial (i.e., before blow-out)
  density profile with $\alpha_i  = 1.0$, whereas  panels on the right
  correspond to $\alpha_i  =  1.5$. In  the upper panels  the baryonic
  mass that is expelled from the galaxy is  assumed to be equal to the
  dark  matter mass that is  relocated to larger  radii (i.e., $\eta =
  1.0$).   In the lower panels we  have assumed that feedback is twice
  as efficient (i.e., $\eta =  2.0$).  The hatched areas correspond to
  the 68.3  percent  confidence  regions   of NGC~247,  DDO~154,   and
  NGC~3109, as labeled in the upper  right panel.  See  the text for a
  discussion.\label{fig:alpha}}


\ifsubmode
\figcaption{\figcapsball}
\figcaption{\figcapvelall}
\figcaption{\figcapchia}
\figcaption{\figcapchib}
\figcaption{\figcapmodels}
\figcaption{\figcaphighres}
\figcaption{\figcapchic}
\figcaption{\figcapfb}
\clearpage
\else\printfigtrue\fi

\ifprintfig

\clearpage
\begin{figure}
\epsfxsize=16.0truecm
\centerline{\epsfbox{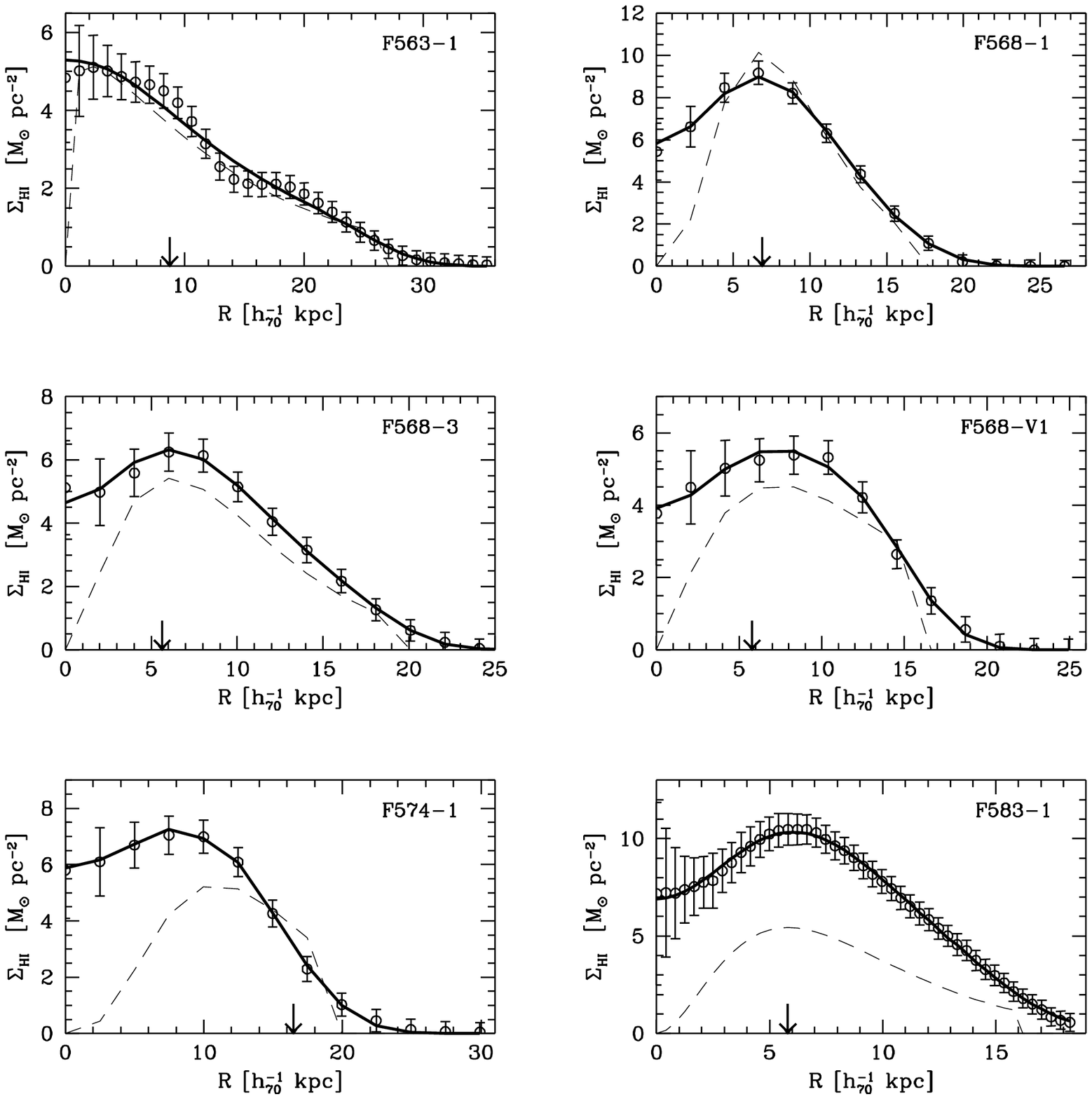}}
\ifsubmode
\vskip3.0truecm
\setcounter{figure}{0}
\addtocounter{figure}{1}
\centerline{Figure~\thefigure}
\else\figcaption{\figcapsball}\fi
\end{figure}

\clearpage
\begin{figure}
\epsfxsize=16.0truecm
\centerline{\epsfbox{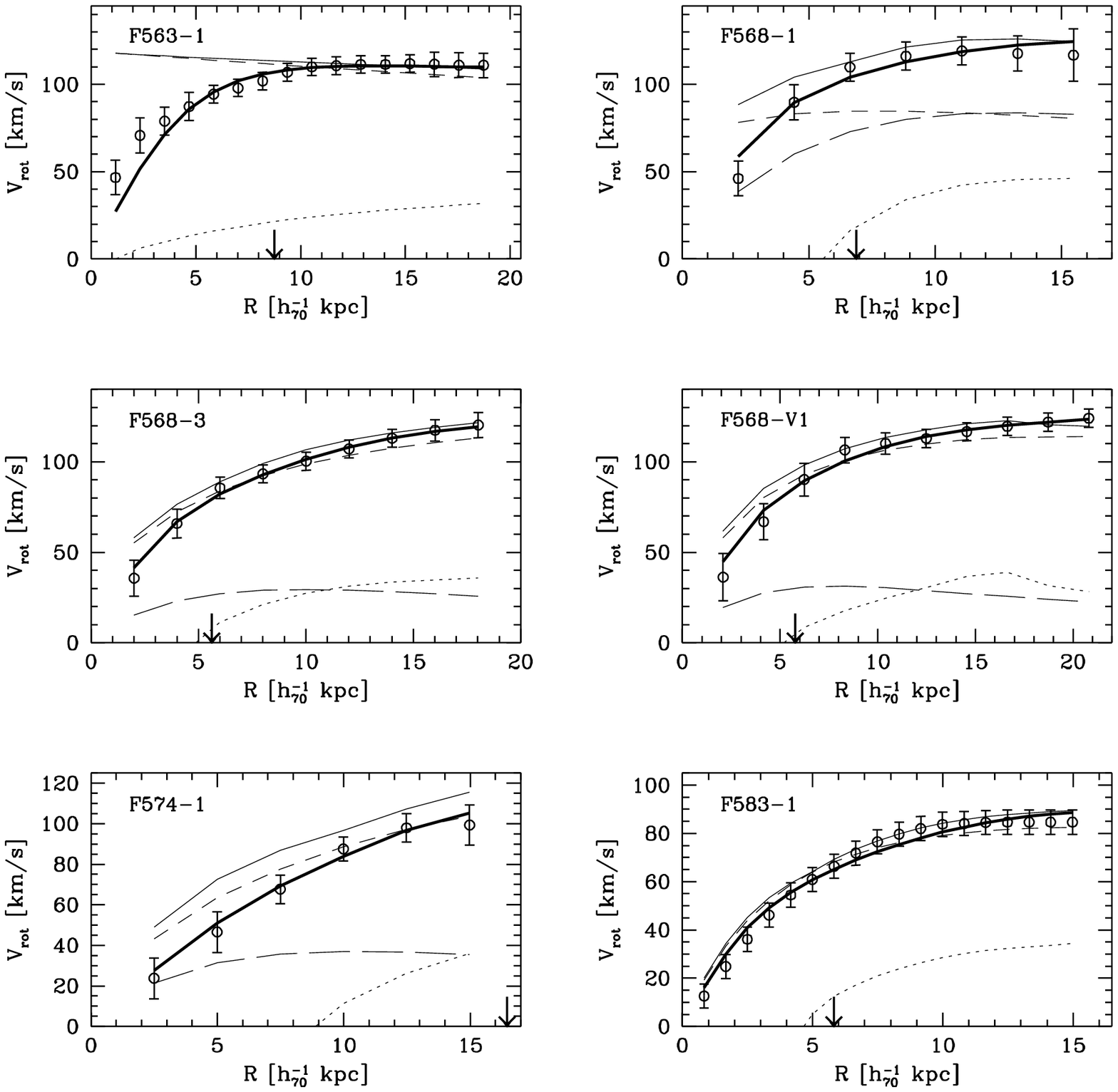}}
\ifsubmode
\vskip3.0truecm
\addtocounter{figure}{1}
\centerline{Figure~\thefigure}
\else\figcaption{\figcapvelall}\fi
\end{figure}

\clearpage
\begin{figure}
\epsfxsize=16.0truecm
\centerline{\epsfbox{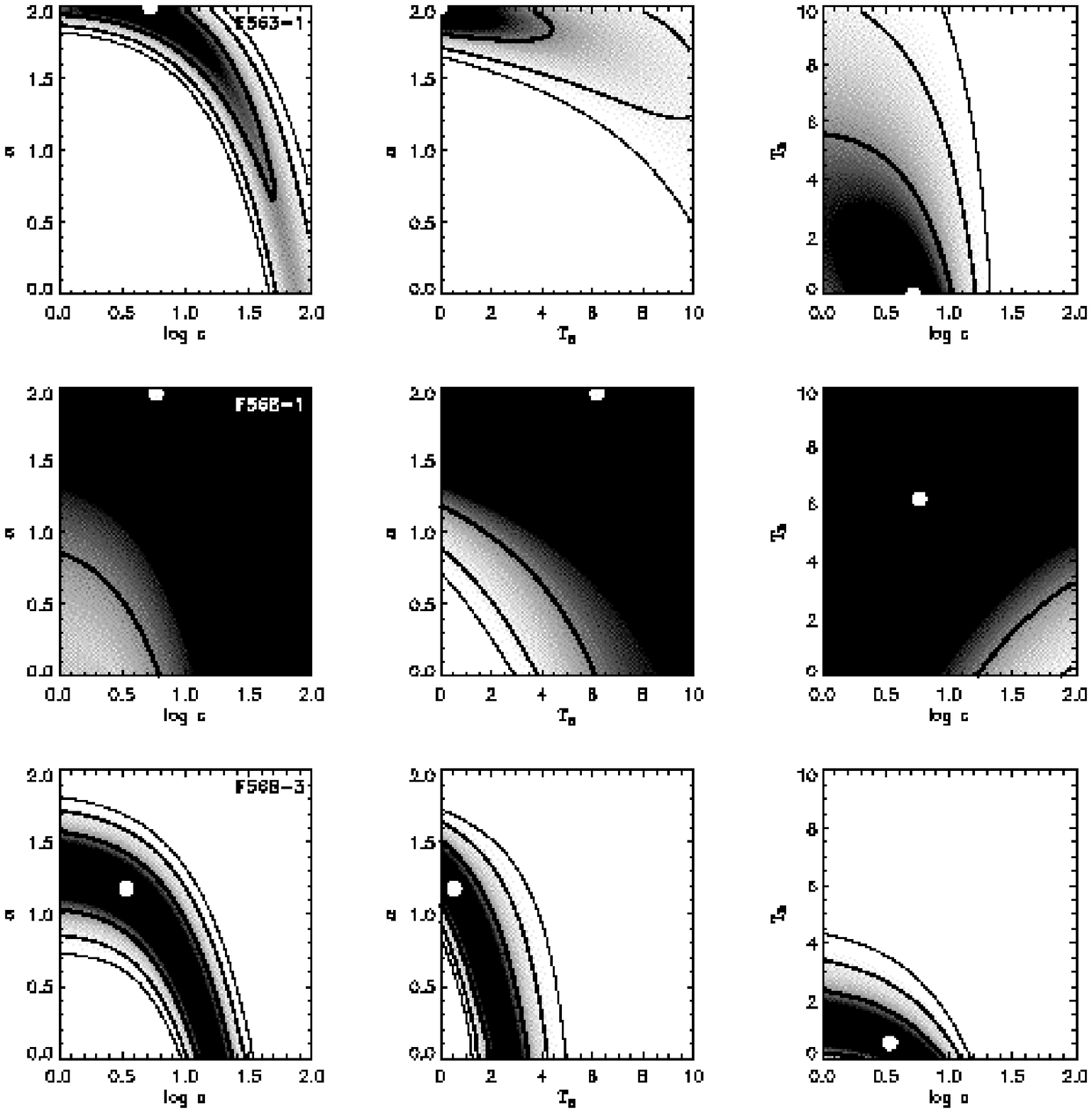}}
\ifsubmode
\vskip3.0truecm
\addtocounter{figure}{1}
\centerline{Figure~\thefigure}
\else\figcaption{\figcapchia}\fi
\end{figure}

\clearpage
\begin{figure}
\epsfxsize=16.0truecm
\centerline{\epsfbox{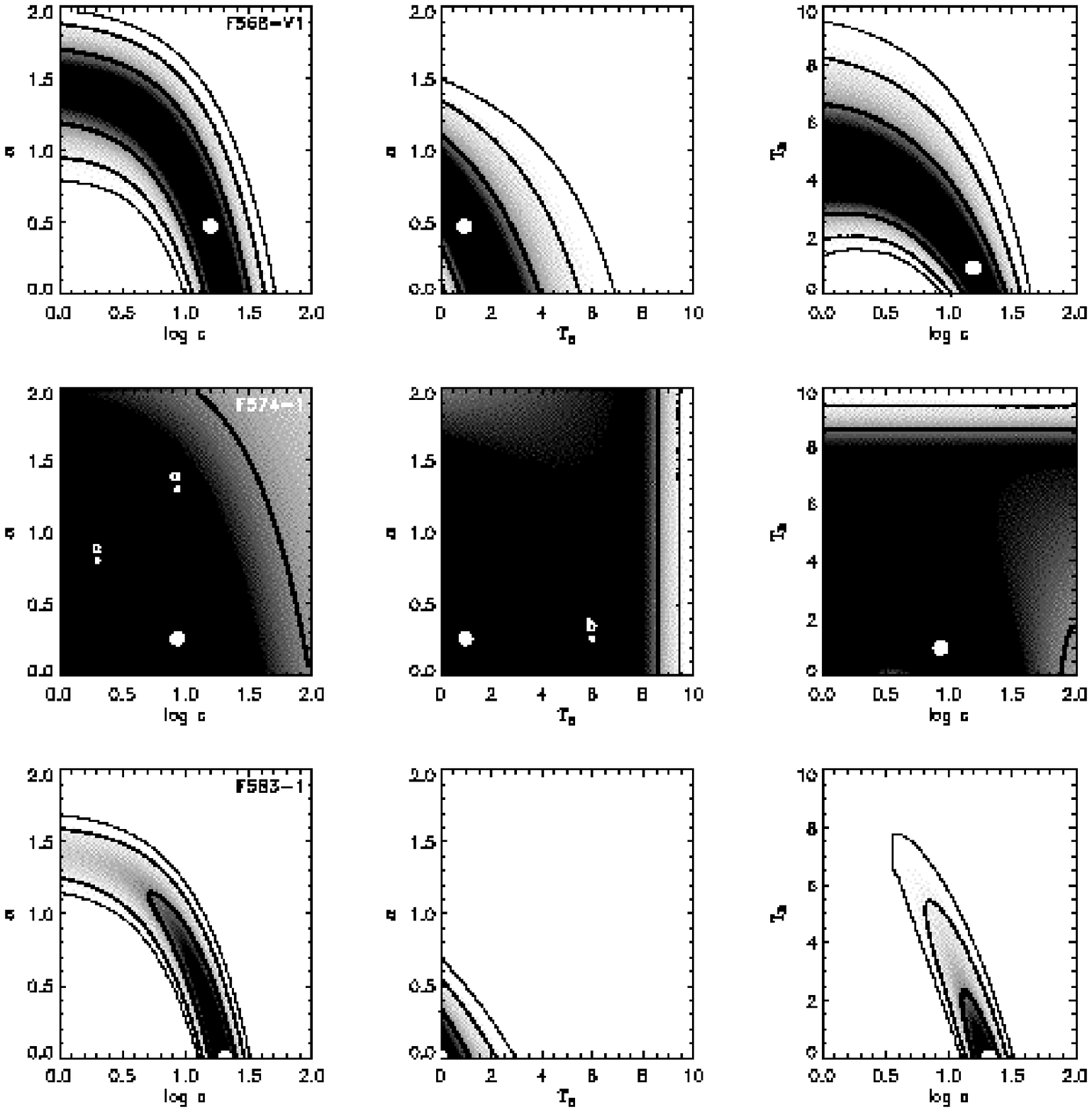}}
\ifsubmode
\vskip3.0truecm
\addtocounter{figure}{1}
\centerline{Figure~\thefigure}
\else\figcaption{\figcapchib}\fi
\end{figure}

\clearpage
\begin{figure}
\epsfxsize=16.0truecm
\centerline{\epsfbox{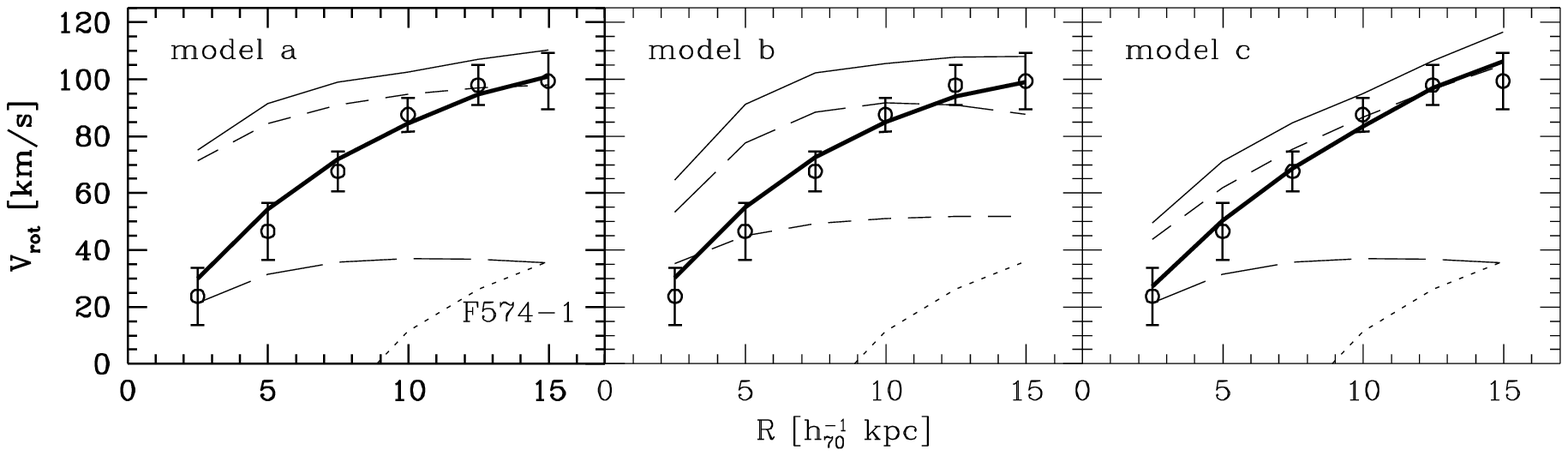}}
\ifsubmode
\vskip3.0truecm
\addtocounter{figure}{1}
\centerline{Figure~\thefigure}
\else\figcaption{\figcapmodels}\fi
\end{figure}

\clearpage
\begin{figure}
\epsfxsize=16.0truecm
\centerline{\epsfbox{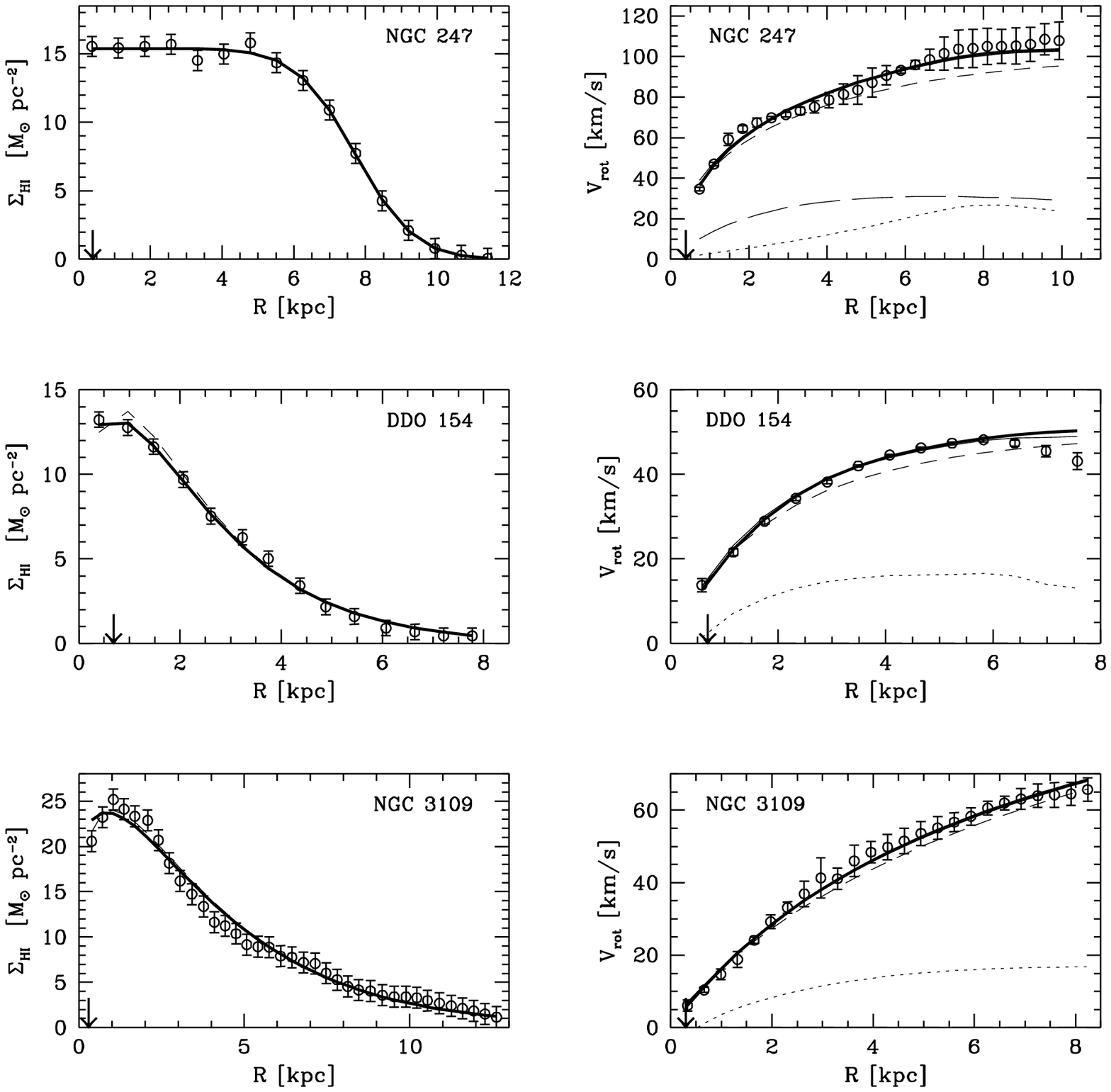}}
\ifsubmode
\vskip3.0truecm
\addtocounter{figure}{1}
\centerline{Figure~\thefigure}
\else\figcaption{\figcaphighres}\fi
\end{figure}

\clearpage
\begin{figure}
\epsfxsize=16.0truecm
\centerline{\epsfbox{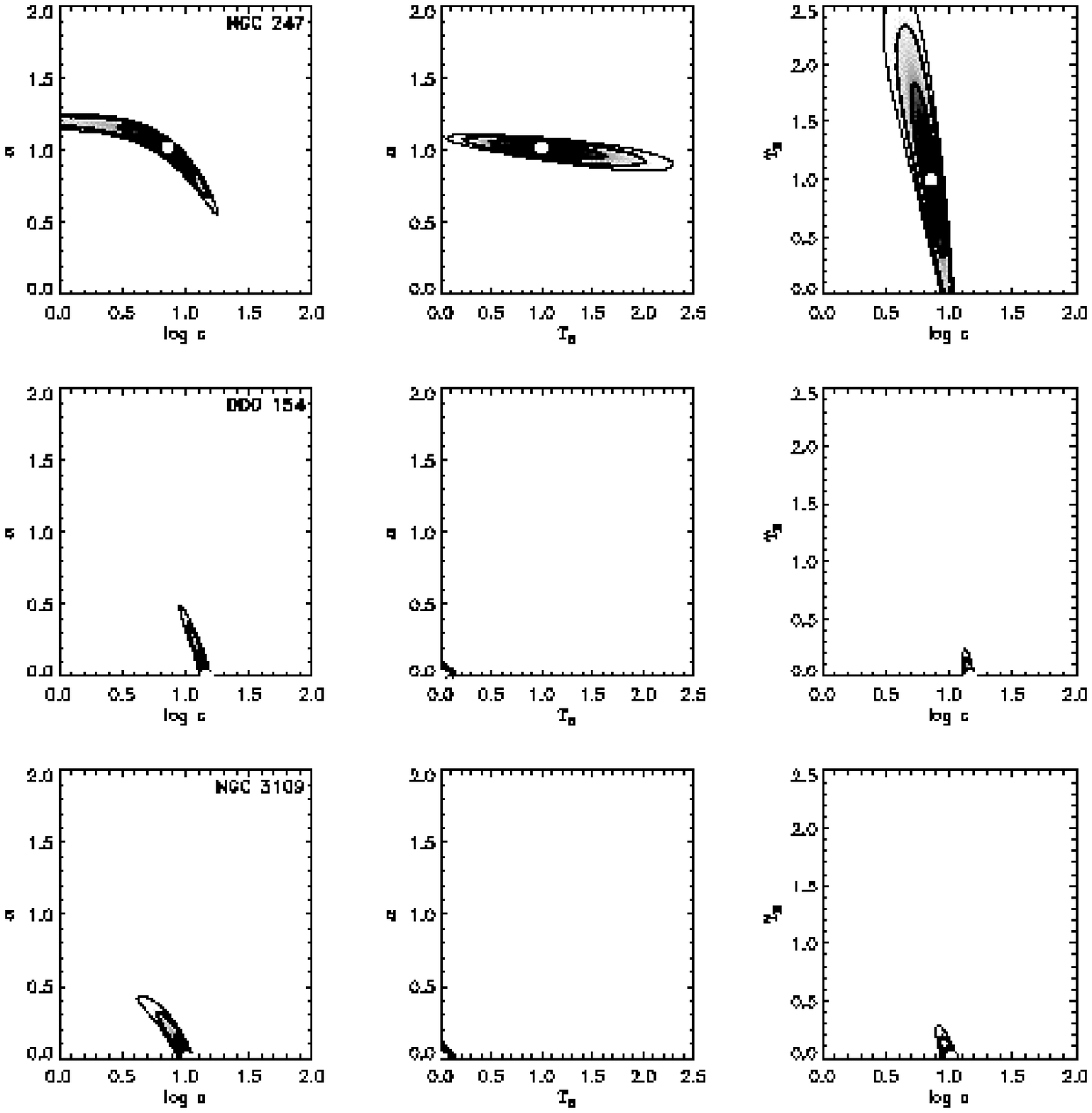}}
\ifsubmode
\vskip3.0truecm
\addtocounter{figure}{1}
\centerline{Figure~\thefigure}
\else\figcaption{\figcapchic}\fi
\end{figure}

\clearpage
\begin{figure}
\epsfxsize=16.0truecm
\centerline{\epsfbox{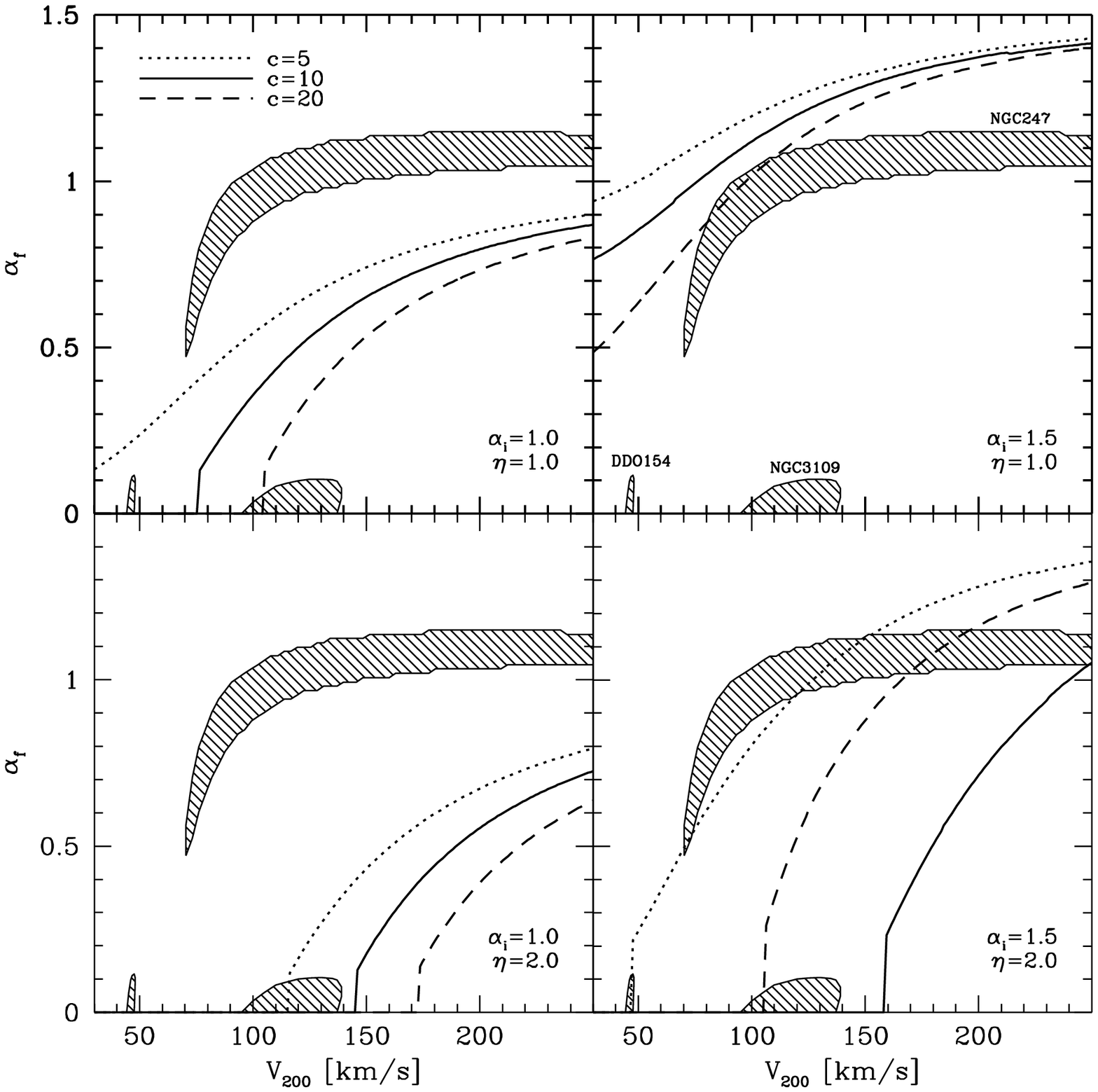}}
\ifsubmode
\vskip3.0truecm
\addtocounter{figure}{1}
\centerline{Figure~\thefigure}
\else\figcaption{\figcapfb}\fi
\end{figure}
\fi



\clearpage
\ifsubmode\pagestyle{empty}\fi


\begin{deluxetable}{lrccrrcrccr}
\tablecaption{Properties of sample of LSB galaxies.\label{tab:data}}
\tablehead{
\colhead{Galaxy} & \colhead{$D$} & \colhead{$M_B$} &
\colhead{$\mu_0^B$} & \colhead{$R_d$} & \colhead{$V_{\rm max}$} & 
\colhead{${\cal R}_1$} & \colhead{${\cal R}_2$} &
\colhead{$i$} & \colhead{$q$} & \colhead{Ref.} \\
\colhead{(1)} & \colhead{(2)} & \colhead{(3)} & 
\colhead{(4)} & \colhead{(5)} & \colhead{(6)} &
\colhead{(7)} & \colhead{(8)} & \colhead{(9)} & 
\colhead{(10)} & \colhead{(11)} \\
}
\startdata
NGC 247   &  $2.5$ & $-17.55$ & $23.23$ & $2.9$ & $108$ & $7.39$ & $24.85$ & $74$ & $0.63$ & 1,2 \\
UGC 11820 & $19.3$ & $-16.92$ & $23.35$ & $2.2$ &  $84$ & $0.54$ &  $4.17$ & $45$ & $0.80$ & 3,4 \\
UGC 634   & $25.0$ & $-17.20$ & $23.14$ & $2.2$ & $107$ & $0.43$ &  $3.61$ & $37$ & $0.88$ & 3,4 \\
F583-1    & $34.3$ & $-16.67$ & $24.03$ & $1.7$ &  $85$ & $0.29$ &  $2.56$ & $63$ & $0.37$ & 7,8 \\
UGC 5999  & $48.2$ & $-17.75$ & $23.50$ & $4.6$ & $155$ & $0.82$ &  $2.90$ & $55$ & $0.86$ & 5,6 \\
F563-1    & $48.6$ & $-17.31$ & $23.64$ & $3.0$ & $111$ & $0.34$ &  $2.14$ & $25$ & $0.34$ & 7,8 \\
F571-8    & $51.4$ & $-17.77$ & $23.87$ & $5.6$ & $133$ & $0.57$ &  $1.71$ & $90$ & $0.34$ & 7,8 \\
UGC 1230  & $54.6$ & $-18.45$ & $23.38$ & $4.8$ & $102$ & $0.76$ &  $5.87$ & $22$ & $0.98$ & 5,6 \\
UGC 5005  & $55.7$ & $-17.95$ & $23.80$ & $4.7$ &  $99$ & $0.71$ &  $4.52$ & $41$ & $0.95$ & 5,6 \\
UGC 128   & $64.3$ & $-18.95$ & $24.23$ & $7.3$ & $131$ & $0.94$ &  $5.86$ & $57$ & $0.94$ & 5,6 \\
F568-3    & $82.9$ & $-18.47$ & $23.08$ & $4.3$ & $120$ & $0.76$ &  $3.20$ & $40$ & $0.92$ & 7,8 \\
F568-V1   & $85.7$ & $-18.07$ & $23.30$ & $3.4$ & $124$ & $0.59$ &  $3.60$ & $40$ & $0.96$ & 7,8 \\
F568-1    & $91.4$ & $-18.27$ & $23.77$ & $5.7$ & $119$ & $0.83$ &  $2.25$ & $26$ & $0.88$ & 7,8 \\
UGC 6614  & $95.7$ & $-21.39$ & $24.28$ &$17.2$ & $228$ & $1.73$ &  $5.84$ & $34$ & $0.93$ & 9,10 \\
F574-1    &$102.9$ & $-18.57$ & $23.31$ & $4.6$ & $100$ & $0.28$ &  $0.91$ & $65$ & $0.39$ & 7,8 \\
F568-6    &$201.4$ & $-22.61$ & $23.38$ &$22.5$ & $320$ & $1.18$ &  $5.64$ & $35$ & $0.94$ & 9,10 \\
          &        &          &         &       &       &        &         &      &        &     \\
DDO~154   &  $4.0$ & $-13.81$ & $23.17$ & $0.5$ &  $48$ & $0.73$ & $11.05$ & $64$ & $0.95$ & 11 \\
NGC~3109  &  $1.7$ & $-16.35$ & $23.17$ & $1.6$ &  $66$ & $5.22$ & $27.78$ & $80$ & $0.75$ & 12,13 \\
\enddata

\tablecomments{Column~(1) lists the name of  the galaxy.   Columns~(2)
  --  (6) list the distance   to  the galaxy (in $h_{70}^{-1}$   Mpc),
  absolute $B$-band magnitude, central $B$-band surface brightness (in
  mag   arcsec$^{-2}$),   scalelength   of   the stellar     disk  (in
  $h_{70}^{-1}$ kpc),   and the   maximum observed rotation   velocity
  $V_{\rm max}$ (in  $\kms$), respectively.  Columns~(7) and~(8)  list
  the ratios of $R_d/S$ and  $R_{\rm max}/S$, with  $S$ the major axis
  of the FWHM  of the beam  of the HI  observations.  These ratios are
  measures for the spatial resolution of the observations. Columns~(9)
  and~(10)    list the  inclination angle    of   the galaxy  and  the
  flattening,  $q$,  of the   beam.   Finally,  column~(11) lists  the
  references to  the original sources of  the  data (see  below).  All
  data has   been converted to  $H_0 =   70  \kmsmpc$.  Magnitudes and
  central surface brightnesses have been corrected for inclination and
  galactic extinction, but not for internal extinction.\\ 1.  Carignan
  \& Puche (1990b), 2.  Carignan (1985b), 3. van Zee \etal (1997),
  4. de Vaucouleurs \etal (1991) 5.  van der Hulst \etal (1993)
  6. de Blok \& McGaugh (1997) 7.  de  Blok, McGaugh \& van der
  Hulst (1996) 8.  de Blok, van  der Hulst \&  Bothun (1995) 9. 
  Pickering  \etal  (1997) 10.    McGaugh \& Bothun  (1994)  11. 
  Carignan \& Beaulieu (1989) 12. Jobin \& Carignan  (1990) 13. 
  Carignan (1985a) \\ }
\end{deluxetable}

\ifemulate\else
  \clearpage
\fi


\begin{deluxetable}{lrrccc}
\tablecaption{Parameters of best fits to HI surface brightness.\label{tab:sbfits}}
\tablehead{
\colhead{Galaxy} & \colhead{$\Sigma_0$} & 
\colhead{$R_d$} & \colhead{$\beta$} &
\colhead{$R_c$} & \colhead{${\rm log}(M_{\rm HI})$} \\
\colhead{} & \colhead{$\Msun {\rm pc}^{-2}$} & 
\colhead{$h_{70}^{-1}$ kpc} & \colhead{} &
\colhead{$h_{70}^{-1}$ kpc} & \colhead{$h_{70}^{-2} \Msun$} \\ 
\colhead{(1)} & \colhead{(2)} & \colhead{(3)} & 
\colhead{(4)} & \colhead{(5)} & \colhead{(6)} \\
}
\startdata
F563-1   &  $8.59$ & $10.63$ & $0.20$ & $26.37$ &  $9.644$ \\
F568-1   &  $4.55$ &  $1.97$ & $3.43$ & $16.98$ &  $9.674$ \\  
F568-3   & $11.52$ &  $3.46$ & $1.78$ & $19.45$ &  $9.524$ \\
F568-V1  & $11.55$ &  $5.24$ & $1.39$ & $15.91$ &  $9.464$ \\  
F574-1   &  $2.16$ &  $3.13$ & $3.51$ & $18.71$ &  $9.649$ \\  
F583-1   &  $9.38$ &  $2.77$ & $2.09$ & $16.18$ &  $9.401$ \\  
NGC~247  &  $4.24$ &  $0.56$ & $7.89$ &  $8.63$ &  $8.912$ \\  
DDO~154  & $14.38$ &  $1.53$ & $0.52$ &  $6.17$ &  $8.383$ \\  
NGC~3109 &  $8.28$ &  $3.08$ & $0.32$ & $12.92$ &  $8.713$ \\  
\enddata

\tablecomments{Column~(1) lists the  name of the  galaxy.  Columns~(2)
  through~(5)  list the best  fitting parameters  for  the HI  surface
  density, and column~(6) lists the  corresponding  HI mass.}
\end{deluxetable}

\ifemulate\else
  \clearpage
\fi


\begin{deluxetable}{lccrrcc}
\tablecaption{Parameters of fits to rotation curves.\label{tab:models}}
\tablehead{
\colhead{Galaxy} & \colhead{Model} & 
\colhead{$\alpha$} & \colhead{$c$} & 
\colhead{$V_{200}$} & \colhead{$\Upsilon_B$} & 
\colhead{$f_{\rm bar}$} \\
\colhead{(1)} & \colhead{(2)} & \colhead{(3)} & 
\colhead{(4)} & \colhead{(5)} & \colhead{(6)} &
\colhead{(7)} \\
}
\startdata
F563-1   & BF & $2.00$ &  $5.2$ &  $73.5$ & $0.0$ & $0.039$ \\
F568-1   & BF & $1.97$ &  $5.8$ &  $64.0$ & $6.2$ & $0.369$ \\
F568-3   & BF & $1.18$ &  $3.4$ & $127.7$ & $0.5$ & $0.010$ \\
F568-V1  & BF & $0.47$ & $15.6$ &  $91.6$ & $0.9$ & $0.023$ \\
F574-1   & BF & $0.26$ &  $8.6$ & $118.3$ & $1.0$ & $0.018$ \\
         &  a & $1.30$ &  $8.6$ &  $76.4$ & $1.0$ & $0.067$ \\
         &  b & $0.26$ &  $8.6$ &  $55.7$ & $6.0$ & $0.537$ \\
         &  c & $0.80$ &  $2.0$ & $278.8$ & $1.0$ & $0.001$ \\
F583-1   & BF & $0.00$ & $20.6$ &  $65.7$ & $0.0$ & $0.035$ \\
NGC~247  & BF & $1.02$ &  $7.2$ &  $93.1$ & $1.0$ & $0.011$ \\
DDO~154  & BF & $0.00$ & $14.7$ &  $44.0$ & $0.0$ & $0.011$ \\
NGC~3109 & BF & $0.00$ & $10.2$ & $101.6$ & $0.0$ & $0.002$ \\
\enddata

\tablecomments{Column~(1)  lists the name  of the galaxy.  Columns~(2)
  lists the  ID of the model, with  `BF' indicating the best-fit model
  (i.e., the one that minimizes $\chi^2_{\rm vel}$).  For F574-1 three
  additional models are listed (a, b, and c) all  of which fall within
  the  68.3 confidence level  of  the BF-model  (see contour plots  in
  Figure~\ref{fig:chib}).  Columns~(3) through~(5) list parameters  of
  the model:   $c$,   $\Upsilon_B$  (in $h_{70}   \Msun/\Lsun$),   and
  $V_{200}$   (in $\kms$). Finally,    column~(7) gives  the resulting
  baryon  fraction    $f_{\rm   bar}   =   (M_{\rm   gas}   +   M_{\rm
    stars})/M_{200}$}
\end{deluxetable}

\clearpage


\end{document}